\documentclass[superscriptaddress,prl,twocolumn,amsmath,amssymb]{revtex4}
\usepackage{graphicx,afterpage}
\usepackage{txfonts}
\usepackage{dcolumn}
\usepackage{bm}
\usepackage{array}
\begin{document}
\preprint{APS/123-QED}
\title{Exotic Landau Diamagnetism and Weyl-Fermions Excitations in TaAs Revealed by $^{75}$As NMR and NQR}
\author{C. G. Wang}
\affiliation{Institute of Physics, Chinese Academy of Sciences,\\
	and Beijing National Laboratory for Condensed Matter Physics,Beijing 100190, China}
\affiliation{School of Physical Sciences, University of Chinese Academy of Sciences, Beijing 100190, China}

\author{Yoshiaki Honjo}
 \affiliation{Department of Physics, Okayama University, Okayama 700-8530, Japan}

\author{L. X. Zhao}
\affiliation{Institute of Physics, Chinese Academy of Sciences,\\
	and Beijing National Laboratory for Condensed Matter Physics,Beijing 100190, China}
\affiliation{School of Physical Sciences, University of Chinese Academy of Sciences, Beijing 100190, China}



\author{G. F. Chen}
\affiliation{Institute of Physics, Chinese Academy of Sciences,\\
 and Beijing National Laboratory for Condensed Matter Physics,Beijing 100190, China}
\affiliation{School of Physical Sciences, University of Chinese Academy of Sciences, Beijing 100190, China}
\author{K. Matano}
 \affiliation{Department of Physics, Okayama University, Okayama 700-8530, Japan}
\author{R. Zhou}
\affiliation{Institute of Physics, Chinese Academy of Sciences,\\
	and Beijing National Laboratory for Condensed Matter Physics,Beijing 100190, China}
\author{Guo-qing Zheng}
\affiliation{Institute of Physics, Chinese Academy of Sciences,\\
 and Beijing National Laboratory for Condensed Matter Physics,Beijing 100190, China}
 \affiliation{Department of Physics, Okayama University, Okayama 700-8530, Japan}
\date{\today}

\begin{abstract}
{The electronic and superconducting properties associated with the topologically non-trivial bands in Weyl semimetals have recently attracted much attention.
We report the microscopic properties of the type-I Weyl semimetal TaAs measured by $^{75}$As nuclear magnetic (quadrupole) resonance under zero and elevated magnetic fields  over a wide temperature range up to 500 K.
The magnetic susceptibility measured by the Knight shift $K$ is found to be negative at low magnetic fields and have a strong field ($B$) dependence as ln$B$ at $T$ = 1.56 K. Such nonlinear field-dependent magnetization can be well accounted for by Landau diamagnetism arising from the 3D linearly dispersed bands, and thus is a fingerprint of topological semimetals. We further study the low-energy excitations by the spin-lattice relaxation rate 1/$T_{1}$. At zero field and 30 K $\leq T\leq$ 250 K, 1/$T_{1}T$ shows a $T^{2}$ variation
due to Weyl nodes excitations.
At  $B \sim$ 13 T,  $1/T_1T$ exhibits the same $T$-dependence but with a smaller value, scaling with $K^2\propto T^2$,  
which   indicates that the Korringa relation also holds for a Weyl semimetal.
Analysis of the Korringa ratio reveals that the energy range of the linear bands is about 250 K in TaAs.
}
\end{abstract}


\maketitle

{

Recently, a new class of topological materials whose low energy excitations are massless chiral fermions, called Weyl semimetals (WSM), has triggered intensive interest in both fundamental physics and applications \cite{WHM,hasan,DingH_PRX,XuSY_science,Fermiarc_STM,DingH_Natphy,LiuZK}. These materials are characterized by the band-touching points known as Weyl nodes around which the spin-nondegenerated bands disperse linearly. The  Fermi arc surface states induced by the Weyl nodes with opposite chirality in the bulk were regarded as a verification of WSM \cite{DingH_PRX,XuSY_science,Fermiarc_STM,Belopolski}. 
However, the Fermi arc states are visible experimentally even in NbP whose Fermi level is outside the energy window of the Weyl cones \cite{XuN}. Therefore, using the observed Fermi arcs as the sufficient condition to identify a Weyl semimetal phase has been challenged \cite{XuN}.
It is important to directly probe the bulk states to identify WSM and detect the Weyl quasi-particles excitations. 
Although a negative magnetoresistivity was observed and attributed to the nonconservation of the particle number with a given chirality induced by non-orthogonal magnetic and electric fields \cite{ChenGF,JiaS}, such interpretation is controversial. Other explanations, such as an experimental artifact resulting from the inhomogeneous exciting current \cite{Reis}, or simply a quantum effect not necessarily related to Weyl bands \cite{Goswami}, were also proposed. 
Therefore, an unambiguous experimental signature from the bulk bands is still lacked.

One alternative attempt to obtain bulk properties is through magnetization measurements. Quantum oscillations of the magnetization have been used to
study the Fermi surface topology \cite{Sergelius,Arnold}. 
Individual Landau Levels due to  linearly dispersed bands in  topological semimetals are essentially different from those in classical free-electron systems \cite{GraphiteLLs,Jeon}, and have a profound effect on the orbital magnetism. As the chemical potential approaches the nodes, a divergent diamagnetism in the weak-field limit will be induced \cite{Koshino,Koshino_1,Thakur,Mikitik}. In particular, the non-oscillating part of the magnetization was suggested to show a nonlinear dependence on the magnetic field beyond a threshold value, through which topological semimetals manifest themselves \cite{nodalline,BitanRoy,Mikitik,Betouras,nodalfermions_Goswami}. Indeed, abnormal field and temperature dependence of the magnetic susceptibility were observed in the 2D Dirac material, graphene \cite{Betouras,LiZL}. However, conventional magnetic measurement failed to detect any nonlinear behavior in the magnetization of 
WSM \cite{TaAs_diamag}.
The nonlinear response effect can be hidden by 
magnetic impurities or disorders in macroscopic magnetic measurements \cite{TaAs_diamag}. 
In addition to the above-mentioned issues, superconductivity derived from topological semimetals has also attracted much attention \cite{MoTe2}. How the bulk Weyl excitations are involved in the superconductivity also becomes an issue.

Nuclear magnetic resonance (NMR) is a local probe suited for addressing the above issues. Firstly, the spin-lattice relaxation rate 1/$T_{1}$ is suited for detecting the excitations in topological materials \cite{Dora,NMR_TaP,Hirosawa}. 
Secondly, the Knight shift $K$ measured by  NMR  can avoid the influence of the impurities and allows us to obtain the intrinsic magnetic susceptibility.
For a conventional metal with trivial bands, the Korringa relation between $1/T_1T$ and $K^2$ has served as a criterion for identifying the nature of metals. It is therefore also important to investigate whether the Korringa relation holds in Weyl semimetals or not.

TaAs is a typical Weyl semimetal whose Weyl nodes are very close to the Fermi level, one set of which is only 1.5 meV away \cite{JiaS}, which is expected to give a more pronounced nonlinear magnetization.
In this work, we use $^{75}$As NMR and NQR (nuclear quadrupole resonance) to get insights into the intrinsic magnetism and Weyl fermion excitations in 
TaAs. We find that the microscopic magnetic susceptibility measured by the Knight shift of $^{75}$As-NMR is negative at low magnetic fields and has a strong field dependence at $T$ = 1.56 K, which can be understood by the exotic Landau diamagnetism peculiar to the topological semimetals. At zero field, the spin-lattice relaxation rate divided by $T$, 1/$T_{1}T$, shows a $T^{2}$ dependence over a wide temperature range 30 K $\leq T\leq$ 250 K,  
due to Weyl nodes excitations. 
At $B$ = 12.95 T, we find a similar $T^{2}$ behavior of 1/$T_{1}T$. The $T$-dependent 1/$T_{1}T$ at the high field is proportional to $K^2$ and indicates that   the Korringa relation  also holds for a Weyl semimetal. Analysis of the Korringa ratio $S$, which is proportional to the squared derivative of $\sqrt{1/T_{1}T}$ with respect to $K$, shows that the low-energy excitations are dominated by Weyl nodes excitations  below $T^*\sim$ 250 K but are due mainly to trivial bands above.


The 
TaAs single crystal samples used in this study were grown by the chemical vapor transport method. The detailed crystal growth method can be found in Ref. \cite{ChenGF}. 
The typical size of the single crystals is 1.5$\times$1$\times$0.5 mm$^{3}$. A single crystal was rotated with respect to the magnetic field direction by using a goniometer in order to have an exact $B \parallel c$-axis configuration \cite{Supple}. The $^{75}$As NMR spectra were obtained by a Fast Fourier Transform (FFT) summation of the spin echo \cite{Clark}. For NQR measurements, a collection of single crystals weighing $\sim$ 1 g was powdered. The $T_1$ was measured by using the saturation-recovery method.

\begin{figure}
\includegraphics[width=7 cm]{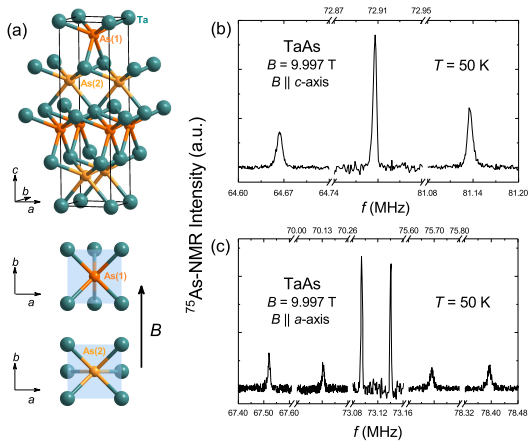}
\centering
\caption{(a) The crystal structure of TaAs (above) and the diagram of two As sites when $B \parallel a$-axis (blow). (b) and (c) are $^{75}$As NMR spectra of TaAs at $T$ = 50 K and $B$ = 9.997 T applied along $c$-axis and $a$-axis, respectively.
}
\label{NMRspectra}
\end{figure}

Figure \ref{NMRspectra} (a) shows the crystal structure of TaAs and Fig. \ref{NMRspectra} (b) and (c) show the $^{75}$As NMR spectra with the magnetic field along the $c$-axis and $a$-axis, respectively. For $B \parallel$ $c$-axis, only three peaks coming from different transitions of the $^{75}$As nucleus with spin $I$ = 3/2 are observed. The full width at half maximum of the central ($\frac{1}{2} \leftrightarrow-\frac{1}{2}$ transition) and satellite ($\pm \frac{3}{2} \leftrightarrow \pm \frac{1}{2}$ transition) peaks are only $\sim$ 3 kHz and $\sim$ 8 kHz at $B$ = 9.997 T, indicating the high sample quality. For $B \parallel a$-axis, two sets of central and satellite lines are observed, which correspond to two types of As sites with different Ta-As-Ta bond direction as shown in Fig. \ref{NMRspectra} (a). The Knight shift with $B \parallel c$-axis, $K_{c}$, is defined as $K_{c}$ = $\frac{f-\gamma B-f_{\text{2nd}}}{\gamma B}$, where $\gamma$ is the gyromagnetic ratio and $f_{\text{2nd}}$ is the second-order quadrupole shift related to 
the electric field gradient (EFG) tensors that can be directly obtained from the spectra shown in Fig. \ref{NMRspectra} \cite{Supple}.

\begin{figure}
\includegraphics[width=8 cm]{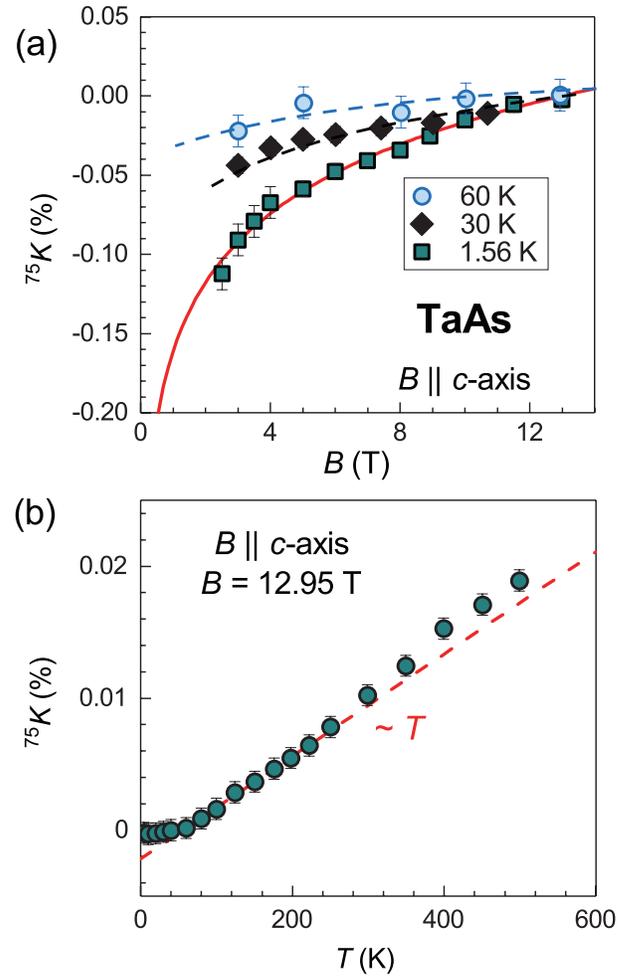}
\centering
\caption{(a) Magnetic field dependence of the $^{75}$As Knight shift $K_{c}$ for ${B} \parallel c$-axis at 1.56 K (squares), 30 K (diamonds) and 60 K (circles). The solid red curve is a fit of $K_{c}$ to $K_{c}$ = $a$ln$B$ + $b$. The obtained parameters are   $a$ = 0.063 with a unit of $\%$$\cdot$ln(T)$^{-1}$  and $b$ = -0.161 $\%$.  The dashed curve is a guide to the eyes. (b) Temperature dependence of the $^{75}$As Knight shift $K_{c}$ for ${B} \parallel c$-axis at 12.95 T. The dashed straight line indicates a $T$-linear relation.
}
\label{Knightshift}
\end{figure}
Figure \ref{Knightshift} (a) shows the obtained field dependent $K_{c}$. At $T$ = 30 K and 60 K, $K_{c}$ is only weakly magnetic field dependent. However, at $T$ = 1.56 K, $K_{c}$ has a very strong field dependence; $K_{c}$ is negative and increases rapidly at low fields. 
The Knight shift measures the uniform magnetic susceptibility $\textrm{$\chi$}$ = $\chi_{\text{spin}}$ + $\chi_{\text{orb}}$ through the relation $K_{c}$ = ${A}_{\text{hf}}^{\text{spin}}\chi_{\text{spin}}$ + ${A}_{\text{hf}}^{\text{orb}}\chi_{\text{orb}}$, where $\chi_{\text{spin}}$  and $\chi_{\text{orb}}$ are the spin and orbital susceptibility respectively. ${A}_{\text{hf}}^{\text{spin}}$ and ${A}_{\text{hf}}^{\text{orb}}$ are spin and orbital hyperfine coupling constants, respectively, which are field independent. For a system with only trivial bands, $\chi_{\text{orb}}$ is the Van-Vleck susceptibility and is both temperature- and field-independent. For a nonmagnetic material, $\chi_{\text{spin}}$ is also field independent, but can have strong field dependence when a small energy gap is closed by magnetic fields, such as superconducting or spin gap \cite{Zhou_PNAS,spingap}, or due to the Kondo effect in heavy fermion systems \cite{heavyfermion}. However, TaAs is a nonmagnetic and weakly correlated system. Therefore,  the field dependence of $K_{c}$
 cannot be due to the  trivial bands which  only contribute a $B$- and $T$-independent value.

The observed field dependence is related to the linear bands. Under magnetic fields, the total energy of the  system will be raised by the electrons from the continuous linear bands that shrink to the zeroth Landau level, leading to a negative $\chi_{\text{orb}}$. Besides, the linear response theory is easily broken for a large magnetic scale (the energy difference between the zeroth and first Landau Level) of the linear dispersed bands even under a moderate magnetic field \cite{Betouras}. Therefore, $\chi_{\text{orb}}$ can have a strong field dependence.
For the case of 2D linear bands, the energy gain $\Delta E$ in the presence of a magnetic field is 
$\Delta E \propto$ $B^{3/2}$. 
Therefore, $\chi_{\text{orb}}$ becomes to be proportional to -1/$\sqrt{{B}}$, which was indeed observed in graphene \cite{LiZL}. 
For 3D linear bands as in TaAs,
an additional quadratic dispersion of the electrons along the direction of the magnetic field has to be considered. 
A logarithmic-like divergence, namely $\chi_{\text{orb}} \propto$ ln${B}$, 
was theoretically suggested 
at the zero-temperature limit when the Weyl nodes are right at the Fermi level \cite{BitanRoy}. We fit the data by $K_{c}$ = $a$ln${B}$+ $b$  and the results are shown in Fig. \ref{Knightshift} (a) by the red solid curve. The agreement between experimental data and theory is good. 
There are two sets of Weyl nodes in this system \cite{JiaS}, and our result implies that at least one set is indeed very close to the Fermi level. 

The nonlinear magnetization is  a result of a failure of the linear response approximation as elaborated below. The linear bands in TaAs with Fermi velocity 10$^{5}$ $\sim$ 10$^{6}$ m/s give a magnetic energy scale to be $v_{F} \sqrt{2 e \hbar B} >$  65 K for $B$ = 2.5 T, being far greater than 1.56 K. This is why the observed nonlinear behavior of the magnetization is strong. Indeed, the field dependence of $K_{c}$ is strongly suppressed at $T$ = 60 K and 30 K compared to that at $T$ = 1.56 K, as can be seen in Fig. \ref{Knightshift} (a),  due to the thermal broadening of the electron distribution
so that a portion of the electrons are excited to higher Landau levels. 
Such behaviors were  not observed by the bulk magnetic susceptibility measurement in previous studies \cite{TaAs_diamag}, probably hidden by impurities inside the sample. Therefore, our results demonstrate that a local probe is indeed crucial for studying this problem.

Figure \ref{Knightshift} (b) shows the temperature dependence of $K_{c}$ at $B$ = 12.95 T. 
  $K_{c}$ increases linearly for 60 K $\leq T \leq$ 250 K, which can be understood as due to the diffuse of electron distribution in the Landau levels at a finite temperature.
  $K_{c}$  at $B$= 8.9 and 6 T shows a similar $T$-linear behavior in this temperature range (see Supplementary Materials \cite{Supple}).
  In the 2D Dirac material, graphene, it was shown that the magnetic susceptibility follows a $T$-linear variation \cite{LiZL}. To our knowledge, our result is the first observation of a $T$-linear magnetic susceptibility in a 3D Weyl semimetal.
  Above $T$ = 250 K, however, $K_{c}$ deviates from the $T$-linear behavior and shows an upturn.
  Our result thus indicates that the linear bands have a finite energy range $E^{*}/k_B\sim$250 K,  and  at high temperatures the trivial bands are involved to contribute to $K_{c}$ when partial electrons are excited to high energies beyond $E^{*}$ where the bands are no longer linearly dispersed.
  For TaAs, first-principle calculations \cite{XuN,JiaS} suggested that $E^{*}/k_B$ is in the range of $\sim$350 K, but has not been decided before by experiments.

At low temperatures below 30 K, on the other hand, $K_{c}$ is almost $T$-independent. This is consistent with the fact inferred from our data that at least one set of the Weyl points is very close to 
the Fermi level, which acts against the thermal disturbance so that $K_{c}$ becomes a constant \cite{LiZL}.
Finally, we note that $K_{c}$ only increases by 0.02$\%$ from 40 K up to 500 K, which is much smaller than the change induced by the magnetic field at 1.56 K. Thus, the large  diamagnetism and its significant field dependence can be viewed as a fingerprint of topological semimetals.

\begin{figure}
\includegraphics[width=7 cm]{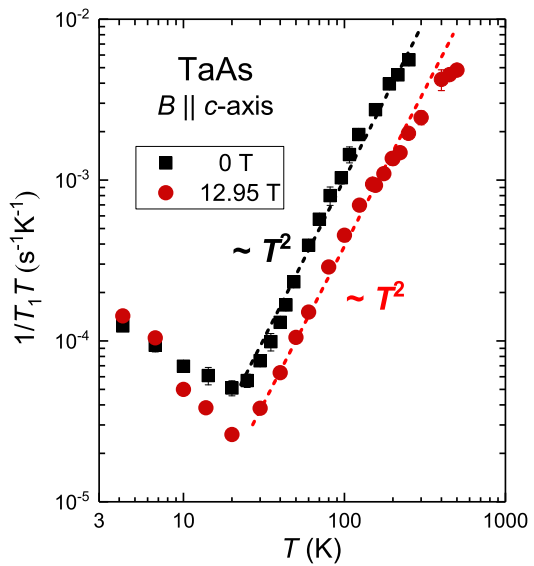}
\centering
\caption{Log-log plot of $^{75}$As $1/T_{1}T$ at 0 T (solid square) and 12.95 T (solid circle). The dashed lines indicate 1/$T_{1}T \propto T^{2}$
}
\label{T1}
\end{figure}

Next, we turn to the spin-lattice relaxation rate to investigate Weyl Fermions excitations.
Figure \ref{T1} shows the comparison of  1/$T_{1}T$ at zero field measured by NQR and at $B$ = 12.95 T by NMR. The details about ($1/T_{1}$)$_{\text {NQR}}$ measurements are documented in the Supplemental Materials  \cite{Supple}.
 Under both fields, upon increasing the temperature, 1/$T_{1}T$ undergoes a crossover from a slow decrease to a rapid increase 
 at $T$ $\sim$ 20 K.
 However, the absolute value is quite different for the two fields.
 Theoretically, ($1/T_{1}$)$_{\text {NMR}}$ and ($1/T_{1}$)$_{\text {NQR}}$ measure the magnetic perturbation perpendicular to the magnetic field and the principle-axis of the EFG, respectively. In the current situation where $B \parallel c$-axis and the principle-axis of EFG is the $c$-axis, they should be the same if there is no additional effect induced by the magnetic field. Therefore, the difference between ($1/T_{1}$)$_{\text {NMR}}$ and ($1/T_{1}$)$_{\text {NQR}}$ originates from other physics than the anisotropy of $1/T_{1}$.

 We start with the zero-field results. 
 Usually, for 3D bands with point nodes, the density of states  (DOS) $N(E)$ $\sim$ $E^{2}$, and 1/$T_{1}$ is proportional to $\int A_{\text{hf}}^{2} N(E)^{2} f(E)[1-f(E)] d E$, where $f(E)$ is the Fermi-Dirac distribution function. Thus 1/$T_{1}T \sim T^{4}$ is expected \cite{Katayama,Yang_PRL,Luo_PRL} when the hyperfine coupling is dominated by spin hyperfine interaction which is a constant. Over a wide temperature range 30 K $\leq T\leq$ 250 K, however, we find that 1/$T_{1}T$ is proportional to $T^{2}$. The same behavior was also reported for 30 K $\leq T\leq$ 100 K by  $^{181}$Ta-NQR  in TaP \cite{NMR_TaP}, where the two sets of Weyl points are quite far away from the Fermi level by 13 meV and 41 meV \cite{Arnold}, respectively, and  thus the system has a smaller $E^*/k_B\sim$150 K \cite{XuN,Arnold}.
Our result can be understood as due to the unusual orbital hyperfine interaction $\mathcal{H}_{\text{hf}}^{\text{orb}}$ for massless Weyl fermions \cite{Dora,Hirosawa}. $\mathcal{H}_{\text{hf}}^{\text{orb}}$ diverges as 1/$q$ where $q$ is the transfer moment of the scattering electrons, leading to $A_{\text{hf}}$ $\propto$ 1/$E$ \cite{Dora,Dora_1}. As a result, 1/$T_{1}T$ behaves as $T^{2}$ instead of $T^{4}$. 
 Combined with the previous study on TaP \cite{NMR_TaP}, the NQR relaxation demonstrates that the 1/$T_{1}T$ $\propto$ $T^{2}$ behavior is indeed an important feature of type-\uppercase\expandafter{\romannumeral1} WSM. 

Below $T_{\text{cross}}\sim$  20 K, 1/$T_{1}T$ starts to increase with decreasing temperature. The upturn of 1/$T_{1}T$ at low temperatures was also observed in TaP by $^{181}$Ta-NQR \cite{NMR_TaP,Supple}.
The crossover temperature $T_{\text{cross}}$ was proposed to be  determined by the distance between $E_{F}$ and the Weyl points, $\Delta$;   the crossover takes place when $k_{B}T$ becomes comparable to $\Delta$ \cite{Dora_1}.
However, $T_{\text{cross}}$ is quite similar for TaAs and TaP   even though $\Delta$
is very different in the two systems, which suggests that the explanation in \cite{Dora_1} is not supported. As mentioned ahead, applying a magnetic field will totally change the electronic state of the Weyl bands hence a discrepancy between 1/$T_{1}T$ at a high field and zero field is expected if 1/$T_{1}T$ is dominated by the linear bands. However, unlike the dramatic change of $K_{c}$ with the field, we find that 1/$T_{1}T$ at the lowest temperature is field independent,  which  suggests that such an upturn is not related to the Weyl excitations; the origin of this upturn merits more theoretical studies in the future.

\begin{figure}
\includegraphics[width=8 cm]{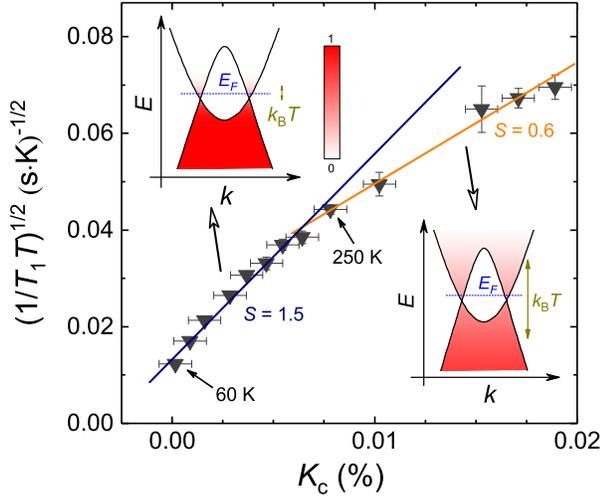}
\centering
\caption{(a) $\sqrt{1 / T_{1} T}$ as a function of $K_{c}$. The two straight lines show the Korringa ratio $S$=1.5 and 0.6, respectively (see text).
	 The insets sketch the  occupancy of the electrons near the Weyl point under zero magnetic field at low temperature (left) and high temperature (right), respectively. The color of shade represents the occupancy probability of electrons. 
}
\label{Korringa}
\end{figure}

The 1/$T_{1}T$ measured at $B$ = 12.95 T  happens to have a $T^{2}$ behavior as well in the temperature range 30 K $\leq T\leq$ 250 K, 
as can be seen in Figure \ref{T1}. 
Note that, in this temperature range, $K_c\propto T$.
The result
can be understood as arising from the fact that at a high field the electrons are reorganized into highly degenerated Landau sub-bands. In such case,  the linear bands are no longer present and the eigenfunction of the Weyl electrons changes. As a result,  the anomalous orbital hyperfine coupling worked at zero-field limit is no longer applicable.
 As discussed already, the motion of the electrons in the quantized cyclotron orbits is the main source of the evolution of the Knight shift with $B$ and $T$. Under a high field, the current fluctuations caused by the electron orbital motion in the Landau sub-bands will create a channel for nuclear spins to relax and contributes to $1/T_{1}T$.
 In light of the $T$-linear behavior of Knight shift shown in Fig. \ref{Knightshift} (b), our observation indicates  $K_{c}^{2} T_{1} T$=const. Namely,  the Korringa relation holds as well for the Weyl excitations.

In order to obtain further insight, we plot $\sqrt{1 / T_{1} T}$ vs $K_{c}$ as shown in Fig. \ref{Korringa}. 
The slop of the $\sqrt{1 / T_{1} T}$ vs $K_{c}$ curve is the so-called Korringa ratio  $S=\hbar\left(\gamma_{e} / \gamma_{n}\right)^{2} \left[d(\sqrt{1/T_{1}T})/d\left(K_{c}\right)\right]^{2}/ 4 \pi k_{B}$.
For a non-correlated metal with trivial bands, theoretically $S$ equals to 1. In  reality, $S$ can deviate from 1. For example, $S\sim$0.6 for simple metals such as Li, Na, and Cs.
As seen in Fig. \ref{Korringa}, in the range 60 K $\leq T \leq$ 250 K where $K\propto T$ and $1/T_1T\propto T^2$, $S \sim$ 1.5. Above $T^{*}=E^*/k_B$$\sim$ 250 K, $S$ crosses over to a smaller value of 0.6. Recall that
 $T^{*}$ is the temperature above which
 1/$T_{1}T$ starts to deviate from the $T^{2}$ behavior and $K_c$ departs  from the $T$-linear line.
 As mentioned already,  this is due to  the linear dispersion bands having a finite energy range $E^{*}$, beyond which  the thermal excitation energy exceeds $E^{*}$.
   At low temperatures, the  excitations are  well within the Weyl energy window (see the sketches in  Fig. \ref{Korringa}).  At high temperatures, on the other hand, the electrons are excited into the trivial bands  and the electron occupancy probability in the linear bands is reduced (diffuse distribution). As a result,   1/$T_{1}T$ and $K_{c}$ are dominated by the trivial bands at high temperatures.
 It is therefore concluded that the Korringa ratio due to the linear bands  is enhanced by a factor 2.5 over that due to the trivial bands in TaAs.
In summary, we have performed the microscopic investigations on the Weyl semimetal TaAs. Remarkably, we observed a clear  ln$B$ dependence of the $^{75}$As Knight shift $K_c$ at $T$ = 1.56 K, which can only be explained by the exotic orbital diamagnetism due to the distinctive Landau levels of the 3D linear bands. Our results demonstrate that such peculiar orbital diamagnetism is a fingerprint of topological semimetals. At zero filed, our observation of the $T^{2}$ variation of 1/$T_{1}T$ over a wide  temperature range 30 K $\leq T\leq$ 250 K is a manifestation of the anomalous orbital hyperfine coupling due to linearly dispersed bands and the $E^{2}$ dependence of the DOS associated with the Weyl excitations.
We further find that,  at $B$ = 12.95 T, 1/$T_{1}T$ follows the same $T$-dependence    but with a smaller value. The scaling between 1/$T_{1}T$ and $K_c^2$ under a high magnetic field indicates that the Korringa relation also holds for a Weyl semimetal. The Korringa ratio due to the linear bands that have an energy range of $\sim$ 250 K is enhanced by a factor of 2.5 over that due to the trivial bands. 


We thank J. Yang, S. Kawasaki and J. Luo for help in some of the measurements, B. Dora, Z. L. Li, K. Nomura, M. Ogata and H. M. Weng  for useful discussion. This work was partially supported by NSFC Grant (No. 2012CB821402) and   MOST Grants (No. 2017YFA0302904 and No. 2016YFA0300502), as well as JSPS grant (No.JP19H00657 and No.JP15H05852).

\clearpage


\vspace{0.5cm}

\begin{references}









\bibitem{WHM}

H. Weng, C. Fang, Z. Fang, B. A. Bernevig, and X. Dai, Weyl Semimetal Phase in Noncentrosymmetric Transition-Metal Monophosphides, Phys. Rev. X \textbf{5}, 011029 (2015).

\bibitem{hasan}

S. M. Huang, S. Y. Xu, I. Belopolski, C. C. Lee, G. Q. Chang, B. K. Wang, N. Alidoust, G. Bian, M. Neupane, C. L. Zhang, S. Jia, A. Bansil, H. Lin, and M. Z. Hasan, A Weyl Fermion semimetal with surface Fermi arcs in the transition metal monopnictide TaAs class, Nat. Commun. \textbf{6}, 7373 (2015).


\bibitem{DingH_Natphy}

B. Q. Lv, N. Xu, H. M. Weng, J. Z. Ma, P. Richard, X. C. Huang, L. X. Zhao, G. F. Chen, C. E. Matt, F. Bisti, V. N. Strocov, J. Mesot, Z. Fang, X. Dai, T. Qian, M. Shi, and H. Ding, Observation of Weyl nodes in TaAs, Nat. Phys. \textbf{11}, 724 (2015).

\bibitem{LiuZK}

Z. K. Liu, L. X. Yang, Y. Sun, T. Zhang, H. Peng, H. F. Yang, C. Chen, Y. Zhang, Y. F. Guo, D. Prabhakaran, M. Schmidt, Z. Hussain, S. K. Mo, C. Felser, B. Yan, and Y. L. Chen, Evolution of the Fermi surface of Weyl semimetals in the transition metal pnictide family, Nat. Mater. \textbf{15}, 27 (2016).

\bibitem{DingH_PRX}
B. Q. Lv, H. M. Weng, B. B. Fu, X. P. Wang, H. Miao, J. Ma, P. Richard, X. C. Huang, L. X. Zhao, G. F. Chen, Z. Fang, X. Dai, T. Qian, and H. Ding, Experimental Discovery of Weyl Semimetal TaAs, Phys. Rev. X \textbf{5}, 031013 (2015).


\bibitem{XuSY_science}
S. Y. Xu, I. Belopolski, N. Alidoust, M. Neupane, G. Bian, C. L. Zhang, R. Sankar, G. Q. Chang, Z. J. Yuan, C. C. Lee, S. M. Huang, H. Zheng, J. Ma, D. S. Sanchez, B. K. Wang, A. Bansil, F. C. Chou, P. P. Shibayev, H. Lin, S. Jia, and M. Z. Hasan, Discovery of a Weyl fermion semimetal and topological Fermi arcs, Science \textbf{349}, 613 (2015).

\bibitem{Fermiarc_STM}

H. Inoue, A. Gyenis, Z. J. Wang, J. Li, S. W. Oh, S. Jiang, N. Ni, B. A. Bernevig, and A. Yazdani, Quasiparticle interference of the Fermi arcs and surface-bulk connectivity of a Weyl semimetal, Science \textbf{351}, 1184 (2016).





\bibitem{Belopolski}
I. Belopolski, S. Y. Xu, D. S. Sanchez, G. Q. Chang, C. Guo, M. Neupane, H. Zheng, C. C. Lee, S. M. Huang, G. Bian, N. Alidoust, T. R. Chang, B. K. Wang, X. Zhang, A. Bansil, H. T. Jeng, H. Lin, S. Jia, and M. Z. Hasan, Criteria for Directly Detecting Topological Fermi Arcs in Weyl Semimetals, Phys. Rev. Lett. \textbf{116}, 066802 (2016).

\bibitem{XuN}
N. Xu, G. Autes, C. E. Matt, B. Q. Lv, M. Y. Yao, F. Bisti, V. N. Strocov, D. Gawryluk, E. Pomjakushina, K. Conder, N. C. Plumb, M. Radovic, T. Qian, O. V. Yazyev, J. Mesot, H. Ding, and M. Shi, Distinct Evolutions of Weyl Fermion Quasiparticles and Fermi Arcs with Bulk Band Topology in Weyl Semimetals, Phys. Rev. Lett. \textbf{118}, 106406 (2017).



\bibitem{ChenGF}

X. C. Huang, L. X. Zhao, Y. J. Long, P. P. Wang, D. Chen, Z. H. Yang, H. Liang, M. Q. Xue, H. M. Weng, Z. Fang, X. Dai, and G. F. Chen, Observation of the Chiral-Anomaly-Induced Negative Magnetoresistance in 3D Weyl Semimetal TaAs, Phys. Rev. X \textbf{5}, 031023 (2015).

\bibitem{JiaS}

C. L. Zhang, S. Y. Xu, I. Belopolski, Z. J. Yuan, Z. Q. Lin, B. B. Tong, G. Bian, N. Alidoust, C. C. Lee, S. M. Huang, T. R. Chang, G. Q. Chang, C. H. Hsu, H. T. Jeng, M. Neupane, D. S. Sanchez, H. Zheng, J. F. Wang, H. Lin, C. Zhang, H. Z. Lu, S. Q. Shen, T. Neupert, M. Z. Hasan, and S. Jia, Signatures of the Adler-Bell-Jackiw chiral anomaly in a Weyl fermion semimetal, Nat. Commun. \textbf{7}, 10735 (2016).

\bibitem{Reis}

R. D. Dos Reis, M. O. Ajeesh, N. Kumar, F. Arnold, C. Shekhar, M. Naumann, M. Schmidt, M. Nicklas, and E. Hassinger, On the search for the chiral anomaly in Weyl semimetals: the negative longitudinal magnetoresistance, New J. Phys. \textbf{18}, 085006 (2016).

\bibitem{Goswami}
P. Goswami, J. H. Pixley, and S. Das Sarma, Axial anomaly and longitudinal magnetoresistance of a generic three-dimensional metal, Phys. Rev. B \textbf{92}, 075205 (2015).




\bibitem{Sergelius}

P. Sergelius, J. Gooth, S. Bassler, R. Zierold, C. Wiegand, A. Niemann, H. Reith, C. Shekhar, C. Felser, B. H. Yan, and K. Nielsch, Berry phase and band structure analysis of the Weyl semimetal NbP, Sci. Rep. \textbf{6}, 33859 (2016).

\bibitem{Arnold}

F. Arnold, C. Shekhar, S. C. Wu, Y. Sun, R. D. dos Reis, N. Kumar, M. Naumann, M. O. Ajeesh, M. Schmidt, A. G. Grushin, J. H. Bardarson, M. Baenitz, D. Sokolov, H. Borrmann, M. Nicklas, C. Felser, E. Hassinger, and B. H. Yan, Negative magnetoresistance without well-defined chirality in the Weyl semimetal TaP, Nat. Commun. \textbf{7}, 11615 (2016).



\bibitem{GraphiteLLs}

G. Li and E. Y. Andrei, Observation of Landau levels of Dirac fermions in graphite, Nat. Phys. \textbf{3}, 623 (2007).

\bibitem{Jeon}

S. Jeon, B. B. Zhou, A. Gyenis, B. E. Feldman, I. Kimchi, A. C. Potter, Q. D. Gibson, R. J. Cava, A. Vishwanath, and A. Yazdani, Landau quantization and quasiparticle interference in the three-dimensional Dirac semimetal Cd$_{3}$As$_{2}$, Nat. Mater. \textbf{13}, 851 (2014).












\bibitem{Koshino}

M. Koshino and T. Ando, Anomalous orbital magnetism in Dirac-electron systems: Role of pseudospin paramagnetism, Phys. Rev. B \textbf{81}, 195431 (2010).

\bibitem{Koshino_1}
M. Koshino and I. F. Hizbullah, Magnetic susceptibility in three-dimensional nodal semimetals, Phys. Rev. B \textbf{93}, 045201 (2016).

\bibitem{Thakur}

A. Thakur, K. Sadhukhan, and A. Agarwal, Dynamic current-current susceptibility in three-dimensional Dirac and Weyl semimetals, Phys. Rev. B \textbf{97}, 035403 (2018).

\bibitem{Mikitik}

G. P. Mikitik and Yu. V. Sharlai, Magnetic susceptibility of topological semimetals, J. Low Temp. Phys. \textbf{197}, 272 (2019).

\bibitem{nodalline}

G. P. Mikitik and Y. V. Sharlai, Magnetization of topological line-node semimetals, Phys. Rev. B \textbf{97}, 085122 (2018).

\bibitem{BitanRoy}

B. Roy and J. D. Sau, Magnetic catalysis and axionic charge density wave in Weyl semimetals, Phys. Rev. B \textbf{92}, 125141 (2015).

\bibitem{nodalfermions_Goswami}
A. Ghosal, P. Goswami, and S. Chakravarty, Diamagnetism of nodal fermions, Phys. Rev. B \textbf{75}, 115123 (2007).

\bibitem{Betouras}

S. Slizovskiy and J. J. Betouras, Nonlinear magnetization of graphene, Phys. Rev. B \textbf{86}, 125440 (2012).


\bibitem{LiZL}

Z. Li, L. Chen, S. Meng, L. Guo, J. Huang, Y. Liu, W. Wang, and X. Chen, Field and temperature dependence of intrinsic diamagnetism in graphene: Theory and experiment, Phys. Rev. B \textbf{91}, 094429 (2015).






\bibitem{TaAs_diamag}

Y. Liu, S. Prucnal, S. Q. Zhou, Z. L. Li, L. W. Guo, X. L. Chen, Y. Yuan, F. Liu, and M. Helm, Intrinsic diamagnetism in the Weyl semimetal TaAs, J. Mag. Mag. Mat \textbf{408}, 73 (2016).

\bibitem{MoTe2}
Y. P. Qi, P. G. Naumov, M. N. Ali, C. R. Rajamathi, W. Schnelle, O. Barkalov, M. Hanfland, S. C. Wu, C. Shekhar, Y. Sun, V. Suss, M. Schmidt, U. Schwarz, E. Pippel, P. Werner, R. Hillebrand, T. Forster, E. Kampert, S. Parkin, R. J. Cava, C. Felser, B. H. Yan, and S. A. Medvedev, Superconductivity in Weyl semimetal candidate MoTe$_{2}$, Nat. Commun. \textbf{7}, 11038 (2016).












\bibitem{Dora}
Z. Okv$\acute {\text {a}}$tovity, F. Simon, and B. D$\acute {\text {o}}$ra, Anomalous hyperfine coupling and nuclear magnetic relaxation in Weyl semimetals, Phys. Rev. B \textbf{94}, 245141 (2016).

\bibitem{NMR_TaP}

H. Yasuoka, T. Kubo, Y. Kishimoto, D. Kasinathan, M. Schmidt, B. Yan, Y. Zhang, H. Tou, C. Felser, A. P. Mackenzie, and M. Baenitz, Emergent Weyl Fermion Excitations in TaP Explored by $^{181}$Ta Quadrupole Resonance, Phys. Rev. Lett. \textbf{118}, 236403 (2017).





\bibitem{Hirosawa}

T. Hirosawa, H. Maebashi, and M. Ogata, Nuclear Spin Relaxation Time Due to the Orbital Currents in Dirac Electron Systems, J. Phys. Soc. Jpn. \textbf{86}, 063705 (2017).











\bibitem{Supple}

See supplementary materials for additional information that includes Ref.\cite{NMRtextbook}.

\bibitem{NMRtextbook}

A. Abraham, \emph{Principles of Nuclear Magnetism} (Oxford University Press, Oxford, 1961).


\bibitem{Clark}

W. G. Clark, M. E. Hanson, F. Lefloch, and P. Segransan, Magnetic resonance spectral reconstruction using frequency-shifted and summed Fourier transform processing, Rev. Sci. Instrum. \textbf{66}, 2453 (1995).

\bibitem{Zhou_PNAS}

R. Zhou, M. Hirata, T. Wu, I. Vinograd, H. Mayaffre, S. Kramer, A. P. Reyes, P. L. Kuhns, R. Liang, W. N. Hardy, D. A. Bonn, and M. H. Julien, Spin susceptibility of charge-ordered YBa$_{2}$Cu$_{3}$O$_{y}$ across the upper critical field, Proc. Natl. Acad. Sci. U.S.A. \textbf{114}, 13148 (2017).


\bibitem{spingap}

Z. L. Feng, Z. Li, X. Meng, W. Yi, Y. Wei, J. Zhang, Y. C. Wang, W. Jiang, Z. Liu, S. Y. Li, F. Liu, J. L. Luo, S. L. Li, G.-q. Zheng, Z. Y. Meng, J. W. Mei, and Y. G. Shi, Gapped Spin-1/2 Spinon Excitations in a New Kagome Quantum Spin Liquid Compound Cu$_{3}$Zn(OH)$_{6}$FBr, Chin. Phys. Lett. \textbf{34}, 077502 (2017).


\bibitem{heavyfermion}

G. R. Stewart, Heavy-Fermion Systems, Rev. Mod. Phys. \textbf{56}, 755 (1984).







\bibitem{Katayama}

K. Katayama, S. Kawasaki, M. Nishiyama, H. Sugawara, D. Kikuchi, H. Sato, and G. -q. Zheng, Evidence for Point Nodes in the Superconducting Gap Function
in the Filled Skutterudite Heavy-Fermion Compound PrOs$_{4}$Sb$_{12}$: $^{123}$Sb-NQR Study under Pressure, J. Phys. Soc. Jpn. \textbf{76}, 023701 (2007).

\bibitem{Yang_PRL}
J. Yang, Z. T. Tang, G. H. Cao, and G. -q. Zheng, Ferromagnetic Spin Fluctuation and Unconventional Superconductivity in Rb$_{2}$Cr$_{3}$As$_{3}$ Revealed by $^{75}$As NMR and NQR, Phys. Rev. Lett. \textbf{115}, 147002 (2015).

\bibitem{Luo_PRL}
J. Luo,  J. Yang,   R. Zhou, Q.G.  Mu, T.  Liu, Z.A.  Ren, C.J.  Yi, Y.G.  Shi, and G.-q. Zheng,
Tuning the Distance to a Possible Ferromagnetic Quantum Critical Point in A$_{2}$Cr$_{3}$As$_{3}$,
 Phys. Rev. Lett. \textbf{123}, 047001 (2019).

\bibitem{Dora_1}

Z. Okv$\acute {\text {a}}$tovity, H. Yasuoka, M. Baenitz, F. Simon, and B. D$\acute {\text {o}}$ra, Nuclear spin-lattice relaxation time in TaP and the Knight shift of Weyl semimetals, Phys. Rev. B \textbf{99}, 115107 (2019).










\end{references}
\end{document}